\documentclass[submission]{dmtcs-episciences}
\usepackage[T1]{fontenc}
\usepackage[utf8]{inputenc}
\usepackage{comment}
\usepackage{amssymb}
\usepackage{dsfont}
\usepackage{amsmath}
\usepackage{lineno}
\usepackage{complexity}
\usepackage[round]{natbib}
\usepackage{hyperref}
\usepackage{cleveref}
\usepackage{todonotes}
\usepackage{graphicx}
\newlang{\PSNS}{PolySNSC}
\newlang{\BPMD}{BPMDecision}
\newtheorem{theorem}{Theorem}
\newtheorem{lemma}{Lemma}
\newtheorem{corollary}{Corollary}
\newtheorem{proposition}{Proposition}
\newtheorem{definition}{Definition}
\newtheorem{claim}{Claim}
\author[Choudhury, Gupta, and Tewari]{Neelabjo Shubhashis Choudhury\affiliationmark{1}
\and Chetan Gupta\affiliationmark{2}
\and Raghunath Tewari\affiliationmark{1}}
\title[Path Isolation in Surface-Embedded Graphs]{Path Isolation and Reachability in Surface-Embedded Graphs with Restricted Cycle and Cut Structure}
\affiliation{Indian Institute of Technology Kanpur, India\\
Indian Institute of Technology Roorkee, India}
\begin{document}
\maketitle
\begin{abstract}
We study the reachability problem in directed surface-embedded graphs under two structural restrictions, using isolating weight assignments.

First, we consider graphs whose underlying undirected embeddings contain at most polynomially many simple surface non-separating cycles. We develop a polynomially bounded, logspace-computable skew-symmetric edge-weight assignment with nonzero circulation around every simple cycle. This yields an unambiguous logspace (\UL) upper bound for directed reachability. The construction also gives an $\SPL$ upper bound for bipartite perfect matching in the corresponding undirected graph class.

Second, for \(g\)-genus directed graphs admitting a balanced hierarchical genus decomposition with \(O(\log n)\) crossing edges at every partition, we obtain an unambiguous \(O(\log g \cdot \log n)\)-space bound for reachability. Our approach combines path-isolating weights for lower-genus subgraphs using polynomially bounded FKS hashing refinements at successive merging levels.

\end{abstract}
\section{Introduction}

In the reachability problem, given a graph $ G = (V,E)$, and two vertices $s$ and $t$ in $G$,  we need to find whether there exists a path from $s$ to $t$. Deciding reachability between a pair of vertices is an important problem in complexity theory. Reachability in directed graphs precisely captures the complexity class nondeterministic logspace (\NL). Also, the seminal result of Reingold showed that reachability in an undirected graph characterises the class deterministic logspace (\L)  \cite{undirected}. Whether these two classes are equal is a big open question in complexity theory. Unambiguous computation is a restricted version of nondeterministic computation. We know that in a nondeterministic computation, given any input, the answer is \emph{yes} if there exists at least one accepting computation path and the answer is \emph{no} if all computation paths reject. However, in an unambiguous computation, given any input, there is at most one accepting computation path. The input is accepted if exactly one computation path terminates in yes, and no if all the paths terminate in no. The class $\UL$ was first introduced formally by Buntrock et al. in \cite{unambiguity} and then studied in more detail by Àlvarez and Jenner in \cite{counting}. Since the class $\UL$ lies between $\NL$ and $\L$, resolving $\NL \stackrel{?}{=} \UL$ will be a big intermediate step toward the $\NL$ vs. $\L$ question. In 2000, Reinhardt and Allender took the first step in this direction and proved that these classes are equal in a non-uniform computation model \cite{nondetunambiguous}. In a later work, it was shown that if deterministic linear space functions cannot be computed by circuits of size $2^{\epsilon n}$, then $\NL=\UL$ \cite{detspacefunction}. These results suggest that the classes might be equal unconditionally. Since directed reachability is a complete problem for $\NL$, giving  $\UL$ algorithm for reachability will prove that $\NL = \UL$. Although solving general directed reachability has been elusive so far, a lot of progress has been made in this direction by solving reachability in restricted classes of graphs, for example, using isolating weight assignment to establish an upper bound of $\UL$ for planar graphs, bounded genus graphs, logarithmic genus graphs, $K_{3,3}$ and $K_5$ free graphs, single crossing minor free graphs \cite{dirplanar,bddgenus1,gst_19,dyncomplexity,Arora,minorfree} and proving an $\L$ upper bound for bounded tree-width graphs \cite{bddtreewidth}. For the general case, Kalampally and Tewari showed that the problem can be solved by an unambiguous algorithm that takes $O(\log^2 n)$ space and polynomial time \cite{kalampally}. The result was subsequently improved by Prakriya and Melkebeek by reducing the space bound to $O(\log^{1.5} n)$ \cite{prakriya}.

\emph{Perfect Matching} is one of the problems that is closely related to the reachability problem. In the perfect matching problem, we are given an undirected graph, and the goal is to find a disjoint subset of edges of the graph that covers all the vertices of the graph. First polynomial time algorithm to solve this problem dates back to 1965 by Edmonds \cite{Edmonds}. However, the parallel complexity of this problem is not clear, i.e. whether there exists an efficient parallel algorithm (an $\NC$ algorithm) for the problem. Lovász took the first step towards this question and proved that the problem belongs to the class randomized $(\NC)$. Later, Mulmuley et al. also gave a randomized $(\NC)$ algorithm to solve the problem using the isolation lemma. In particular, they proved that if we can efficiently construct a polynomial-size weight assignment for the edges of a graph such that there exists a unique minimum weight perfect matching in the graph, then we can solve the problem efficiently by an $\NC$ algorithm. Allender et al. improved this result in \cite{detspacefunction} and proved that if we can construct such a weight function in logarithmic space, then the problem will belong to the class $\SPL$, which is a subclass of $\NC$. In the last few years, people have constructed such a weight assignment for various restricted classes of graphs \cite{IsolatingPM,bddgenus1,minorfree,lognPerfectMatching}. Recently, Fenner et al. and Svensson and Tarnawski gave a construction of weight functions of size $O(\log^2 n)$-bits, proving that the problem can be solved deterministically by a quasi-$\NC$ algorithm in bipartite and general graphs, respectively \cite{Fenner2016, Svensson2017}. As we see, if we can efficiently construct a weight assignment that isolates the path and perfect matching, then we can obtain the desired bound for both problems. In most of the previous results, people have constructed a single weight assignment for a class of graphs that isolates a path in the graphs of the class and a perfect matching in the corresponding undirected bipartite graphs. These weight functions satisfy two properties:$(i)$ they should be \ emph {skew-symmetric}, and they should give nonzero weight to all the cycles in the graphs. We will also construct such a weight function for a restricted graph class here.

In this work, we focus on directed graphs that can be embedded on surfaces. A cycle in a directed graph refers to a \emph{directed cycle} and one in an undirected graph refers to an \emph{undirected cycle}. A surface has genus $g$ if it can be obtained by attaching $g$ handles to a sphere. The \emph{genus of a graph} is the minimum genus surface on which the graph can be embedded without any edge crossings (For example, planar graphs can be embedded on a surface of genus $0$, i.e. a sphere, $K_5$ can be embedded on a surface of genus $1$). It is easy to see that any graph can be embedded on a surface of polynomially ordered (in the number of vertices) genus. Therefore, solving the problem for such kind of graphs has been another direction for making progress towards this question. If we consider the underlying undirected version of directed graphs embedded on surfaces, then the cycles in such a graph can be partitioned into two classes - {\em surface separating cycles} (SSC) and {\em surface non-separating cycles} (SNSC). It has been proved that for graphs that contain an arbitrary number of surface separating cycles and up to $O(\log n)$ many topologically different surface non-separating cycles, both reachability and perfect matching can be isolated efficiently. The next step is to handle an arbitrary number of surface non-separating cycles. In this paper, we construct path-isolating as well as bipartite perfect matching-isolating weight assignments for a graph that has polynomial number of surface non-separating cycles.

We also investigate a complementary restriction for embedded directed graphs of genus $g$: a hierarchical partition of the surface into lower-genus pieces with only $O(\log n)$ crossing edges at every partition. Even when the pieces have complicated internal topology, the limited number of crossing edges bounds the number of competing minimum-weight paths after their weights have been isolated. An FKS hashing refinement at each merging level leads to an unambiguous $O(\log g\cdot\log n)$-space bound. Thus, the two contributions use path isolation to handle different forms of structural restriction: the number of surface non-separating cycles and the width of genus partitions.

\subsection{Our Results}

We study two complementary families of surface-embedded graphs. For graphs with polynomially many surface non-separating cycles, we obtain the following reachability and bipartite perfect-matching bounds.

\begin{theorem}
\label{thm:pathmain}
In directed graphs embedded on surfaces that contain at most a polynomial number of surface non-separating cycles, the reachability problem can be solved in unambiguous logspace $(\UL)$.    
\end{theorem}

We complement this result by establishing an analogous complexity bound for the bipartite perfect matching problem.

\begin{theorem}
\label{thm:matchingmain}
In undirected bipartite graphs embedded on surfaces that contain at most a polynomial number of surface non-separating cycles, the bipartite perfect matching problem lies in $\SPL$.
\end{theorem}

For $g$-genus graphs admitting a balanced hierarchical decomposition with logarithmically many crossing edges at every internal partition, we obtain a further reachability bound.

\begin{theorem}
\label{thm:crossmain}
In directed graphs embedded on surfaces that contain $O(\log n)$ crossing edges between any $2$ partitions of its genus decomposition, the reachability problem can be solved unambiguously in $O \bigl( \log g \cdot \log n \bigr)$ space.
\end{theorem}

\paragraph{Organization.} Section~\ref{Sec_Motivation} motivates the cycle restriction. Section~\ref{Sec_path_matching} establishes the PolySNSC reachability result, followed immediately by the bipartite perfect-matching consequence in Section~\ref{Sec_perfectmatch}. The final main section establishes the genus-decomposition reachability bound.

\section{Preliminaries}
Given a directed graph $G=(V,E)$, a \emph{weight assignment} (alternatively function) $w:E \rightarrow \mathbb{Z}$ is a function that assigns integer weights to all the edges of $G$. Consequently, the weight of a path in $G$ is defined as the sum of the weights of edges that form the path. Graph $G$ is said to be \emph{min-unique} with respect to a weight function $w$, if between every pair of vertices, there exists at most one minimum weight path. The weight function $w$ is called a min-isolating (or path-isolating) weight function. A \emph{weight function} $w$ is called \emph{skew-symmetric} if for all edges $e= (u,v) \in E$, $w(e) = - w(e^r)$, where $e^r =(v,u)$. A weight function is called \emph{polynomially bounded} (or of polynomial size) if it assigns $O(\log n)$-bit weights to all the edges of a graph. One of the ways to obtain isolation of paths and perfect matchings in a directed graph is by designing a skew-symmetric weight function that assigns non-zero weights to all the cycles in the graph. 

We know that every graph can be embedded on a \emph{surface} such that no two edges of the graph intersect each other. The \emph{genus of a graph} is defined as the minimum genus surface on which it can be embedded without intersecting its edges. A surface having genus $g$ has $g$ holes in it. Cycles in graphs embedded on a surface of genus $>0$ can be divided into two classes : (i) \emph{surface separating cycles (SSCs)} and, (ii) \emph{surface non-separating cycles (SNSCs)}.

\begin{definition}
Let $G$ be a graph embedded on a surface $\Sigma$. A cycle $C$ is surface-separating if it separates $\Sigma$, i.e., $\Sigma \setminus C$ is disconnected. Otherwise, $C$ is surface non-separating. Alternatively, a cycle in the graph is called surface-separating if removing the cycle from the surface splits the surface into two disconnected regions and it is called non-separating if cutting along the cycle does not disconnect the surface and reduces the genus by 1.
\end{definition}

\begin{figure}
\centering
\includegraphics[width= 0.64 \textwidth]{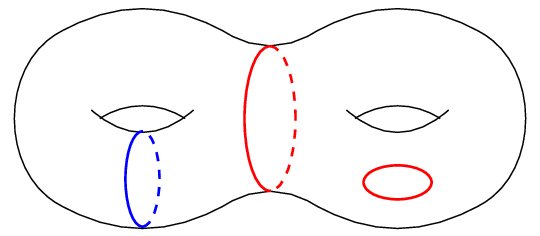}
\caption{Surface separating cycles (marked in red) and surface non-separating cycle (denoted in blue). \href{10.48550/arXiv.2005.01590}{Figure Source}} %\label{fig2}
\label{fig:cycles}
\end{figure}

As the name suggests, a cycle in a surface embedded graph is called a surface non-separating cycle if cutting along the cycle does not divide the surface into two parts and vice versa (see \cref{fig:cycles}). SNSCs are important because they are related to the genus of graphs. The genus of a graph $G$ is at most the number of disjoint SNSCs in $G$, such that the surface does not divide even after cutting it along all the cycles together. In other words, if a graph is embedded on a minimum possible genus surface, then the genus of the graph (and surface) will be equal to the number of SNSCs present in the graph such that cutting the surface along those cycles does not separate the surface. The importance of the class of graphs with poly-many SNSCs is explained in detail in \Cref{Sec_Motivation}. Our objective therefore is to construct a weight function for the above class of graphs.

% \begin{definition}
% A polynomial-size weight function is a weight function that assigns each edge in the graph a weight of $(c \cdot n^k)$ for some constant $c$, where the polynomial bound is on the number of vertices $n$. Such a weight function can be represented using $O(\lg n)$ bits.    
% \end{definition}

\begin{definition}
Let $\PSNS$ be the class of surface-embedded directed graphs such that they contain at most polynomially many surface non-separating cycles.
\end{definition}

For a nondeterministic Turing machine $M$, let $acc_M(x)$ and $rej_M(x)$ denote the number of accepting computations and the number of rejecting computations, respectively, on an input $x$. Denote $gap_M(x) = acc_M(x) - rej_M(x)$. 

\begin{definition} 
A language $L$ is in $\SPL$ if there exists a logspace bounded nondeterministic machine $M$ so that for all inputs $x, gap_M(x) \in \{0,1\}$ and $x \in L$ if and only if $gap_M(x) = 1$.
\end{definition}
Complexity class $\SPL$ was initially studied by Allender et al. in \cite{detspacefunction}.
% F$\L^{\SPL}$ is the class of functions computed by a logspace machine with an $\SPL$ oracle.
%\begin{definition}
With respect to matching, we define the following problem -
$\BPMD$: Given a bipartite graph G, checking if G has a perfect matching \cite{bddgenus1}.
Although the weight function constructed by \cite{gst_19} isolates paths in $O(\log n)$ genus graphs, it is worth noticing that it is a skew-symmetric weight function that gives non-zero weights to all surface separating cycles (but not necessarily to any surface non-separating cycles) in graphs of arbitrary genus. We reuse their weight function as part of our final weight function and encapsulate their result in the following lemma.

\begin{lemma}
\label{lem:ssnonzero}\cite{gst_19}
Let $G$ be a directed graph given along with its embedding on a surface. There exists a logspace-computable polynomial-size skew-symmetric weight function $w_{pl}$ that gives nonzero weight to every surface-separating cycle in $G$.    
\end{lemma} 

We now state the hashing result by Fredman, Koml\'{o}s and Szemer\'{e}di %Fredman, Komlós, and Szemerédi
that we use later. We refer to it as FKS Hashing Lemma in this text. 
\begin{lemma} \cite{fks}
\label{lem:fks}
Let $S = \{x_1, x_2, \ldots, x_k\}$ be a set of n-bit integers. Then there exists an $O(\log n + \log k)$ bit prime number $p$ so that for all $x_i \neq x_j \in S, x_i$ mod $p \neq x_j$ mod $p$.
\end{lemma}

We will use the following Theorem from Reinhardt and Allender \cite{nondetunambiguous} to prove \Cref{thm:pathmain}. 
\begin{theorem} \cite{nondetunambiguous}
\label{thm:ulalgorithm}
There is a nondeterministic logspace Turing machine $M$ that takes $\langle G, s, t,w \rangle$ as input where $G$ is a directed graph on $n$ vertices, $s$ and $t$ are vertices in $G$ and $w$ is an $O(\log n)$-bit edge weight function and outputs the following along a unique computation path while all other computation paths halt and reject :
\begin{itemize}
\item Not Min-unique if G is not min-unique with respect to w,
\item Yes, if G is min-unique with respect to w and there is a path from s to t in G, and
\item No, if G is min-unique with respect to w and there is no path from s to t in G.
\end{itemize}
\end{theorem}

\section{Motivation}
\label{Sec_Motivation}
The class of graphs that contain polynomially many SNSCs is of particular interest due to its intricate relationship with graph embeddings and surface topology. Surface non-separating cycles are closely tied to the genus of a graph (or equivalently, the embedding surface). As the genus increases, more complex graphs can be embedded, and in fact, any graph can be embedded on a surface of polynomial genus. Interestingly, we observe that graphs with poly-many SNSCs can still exhibit very high genus, say of $O(n)$. Conversely, there exist graphs of low genus that nevertheless contain exponentially many SNSCs. This demonstrates that the class $\PSNS$ is neither a strict superset nor a strict subset of the classes for which polynomial-size path-isolating weight functions are currently known. Consequently, identifying such a weight function for this class becomes a compelling and nontrivial direction. Below we give an example of a graph that has a very high genus but only polynomially many SNSCs.

\subsection{Constructing a graph with high genus having poly-many SNSCs}

Let us assume we have $i$ copies of the graph $K_5$, namely graphs $G_1,G_2, \ldots, G_i$, where each copy is embedded separately on a genus $1$ surface (i.e., a torus). Let us assume that $G_i$ contains vertices $\{v,v_1^i,v_2^i,v_3^i,v_4^i\}$. \emph{$1$-clique sum} is a special version of $k$-clique sum that is obtained by gluing two graphs together at a single vertex. This is done by selecting one vertex each from two separate graphs and identifying them as one in the resultant graph, keeping all edges incident to both vertices. Now we repeatedly take $1$-clique sum of all the $G_i$s on $v$. Let $G$ be the final graph thus constructed. We have given an informal proof that the genus of $G$ is $i$ and it contains only polynomially many SNSCs. We will prove it for the case when $i=2$, and the same argument can be extended for arbitrary $i$. We have $G_1$ and $G_2$ each embedded on a torus. We will prove that the genus of $G$ is $2$. For the sake of contradiction, assume that $G$ can be embedded on a surface $S$ of genus $1$.

\begin{definition}
For a graph $G$ embedded in an orientable or non-orientable closed surface $\Sigma$, the embedding is called a \emph{2-cell embedding} or a \emph{cellular embedding} where every face is homeomorphic to an open disk, i.e., each face is connected, and the boundary of each face is a closed walk in the graph. 
\end{definition}

The subgraph of $G$ induced by the vertices of $G_1$ is a non-planar graph, say $H$. The embedding of $H$ on the surface $S$ will be a \emph{2-cell embedding}. No matter how we embed this subgraph on $S$, each face in the embedding will be homeomorphic to an open disc. Notice that the remaining subgraph of $G$, i.e., $(G-H)$, formed by removing only the edges, is also non-planar. Therefore, it cannot be embedded inside any face of $H$, hence contradicting the assumption. This proves that $G$ cannot be embedded on the surface of genus $1$. Similar argument can be extended for arbitrary $i$.

We now show that the number of SNSCs in this graph is linear in $i$. Let us consider a cycle starting from and ending at some $v_p^i \in G_i$ and covering $v$. No vertex $v_p^i$ can form a cycle containing another vertex $v_q^j \in G_j$, since there is only a single way to traverse between them, through $v$. $v$ has to be visited in any path going from $v_p^i$ to $v_q^j$. While completing the cycle back to $v_p^i$, $v$ has to be visited again, as there does not exist any other way to connect them. But a unique vertex cannot be visited twice in a cycle. Hence, each $G_i$ can have at most $12$ SNSCs. If we consider $G$ to be of linear genus, i.e., if $i=n$, then the total number of surface non-separating cycles is at most $12i = O(n)$. Thus $G$ is a graph with high genus (of order linear in $n$), yet having poly-many SNSCs.

\section{Path isolation and reachability in \texorpdfstring{$\PSNS$}{\PSNS}}
\label{Sec_path_matching}

Let us consider a  directed acyclic graph (DAG) $G'=(V_{G'},E_{G'})$  and its corresponding underlying undirected graph, $H'$. Suppose the number of directed SNSCs in $G'$ is $t$. We know that reachability in $G'$ is $\NL$-complete because reachability in a DAG belongs to the class $\NL$ and is also $\NL$-hard. On the other hand, $H'$ can contain exponentially many undirected cycles. Therefore, putting a polynomial bound on the number of directed cycles does not imply a polynomial bound on the number of undirected cycles.

We briefly explain why constructing a skew-symmetric weight function that assigns non-zero weights to all the cycles in a graph is enough to isolate paths in that graph.

\begin{proposition}
\label{prop:skewsymmetric}
For a class of directed graphs, if there exists a skew-symmetric weight function that gives non-zero weight to all the cycles in graphs of that class, then the weight function is path-isolating for that class of directed graphs.
\end{proposition}

\begin{proof}
Let $w_s$ be a skew-symmetric weight function that gives nonzero weights to all the cycles in a directed graph $G$. Let us assume there are two different paths $P_1$ and $P_2$, from $u$ to $v$ that have the same minimum weight. Now there are two cases, either they intersect or they do not intersect. Let us first assume that they do not intersect. Let us construct a graph $G'$, that is obtained from $G$ by only reversing the direction of the edges along path $P_2$. Let us call that path $P_2^r$ in $G'$. Let $C$ be simple cycle formed by paths $P_1$ and $P_2^r$, in $G$. We know that $w_s(C) = w_s(P_1) + w_s(P_2^r)$. Since $w_s$ is skew-symmetric, we have $w_s(P_2) = -w_s(P_2^r)$. This implies that $w_s(C) = w_s(P_1) - w_s(P_2)$. Since $P_1$ and $P_2$ are equal weight we can conclude that $w_s(C) = 0$, which contradicts the assumption. Similarly, when $P_1$ and $P_2$ intersect, we can repeat the argument by considering the two sub-paths of $P_1$ and $P_2$ that start from $u$ and end at their first intersection point.
\end{proof}

\subsection{Deciding reachability in \texorpdfstring{$\PSNS$}{\PSNS}}
\label{Sec_pathreach}

\begin{theorem}
\label{thm:maintheorem}
Let $G=(V,E)$ be a $g$-genus directed graph containing polynomially many simple surface non-separating cycles. Then there exists an $O(\log n)$-bit skew-symmetric weight function $w: E \rightarrow \mathbb{Z}$ computable in logarithmic space that assigns non-zero weights to all the surface non-separating cycles in $G$.
\end{theorem}

\begin{proof}
Let $C = \{C_1, C_2, \cdots, C_t\}$ be the set of $t$ many surface non-separating cycles, $t = O(n^k)$, for some constant $k$. If $C_i$ for any \( i \in [t] \) is not a directed cycle in $G$ then we do not care about it because the goal is to give nonzero weights to all the cycles. Therefore assume each $C_i$ is directed cycle such $C_i$ consists of two paths, namely $P_i$ from $u_i$ to $v_i$ and $P'_i$ from $v_i$ to $u_i$, respectively. 
Let us index all the edges in the graph as $e_1, e_2, \cdots, e_m$. We define a skew-symmetric weight function $w_{exp}$ which assigns $2^i$ to edge $e_i$, i.e, $w_{exp}(e_i) = 2^i$. Weights of paths $P_i$ and $P_i'$ are defined as,
\[
w_{exp}(P_i) = \sum_{\forall e_i \in P_i} 2^i,\qquad w_{exp}(P_i') = \sum_{\forall e_j \in P_i'} 2^j    
\]

Let $P_i'^{r}$ be the path obtained by reversing the edges along path $P_i'$. Let $\mathcal{P} = \{P_1, P_1'^r,P_2, P_2'^r \ldots P_t'^r\}$. We can see that every path $\mathcal{P}$ gets a different weight with respect to $w_{exp}$, but the weight of a path is exponential. Since the cardinality of $\mathcal{P}$ is $O(n^k)$, from \Cref{lem:fks}, we know that there exists an $O(\log n)$-bit prime $p$, such that with respect to the weight function $w_{fks} (e_i)= w_{exp} (e_i)$ mod $p$, each path in $\mathcal{P}$ gets a different weight. $w_{fks}$ here is of polynomial size.
\[
w_{fks} (C_i) = w_{fks} (P_i) -  w_{fks} (P_i'r)    
\]

Notice that, since each path in $\mathcal{P}$ gets a different weight, we can conclude that each $C_i$ get a nonzero weight with respect to $w_{fks}$. Also, from \Cref{lem:ssnonzero}, we know that each surface separating cycle in $G$ gets a nonzero weight with respect to a skew-symmetric weight function $w_{pl}$. Therefore, taking a linear combination of these two weight functions, each cycle in the graph will now have zero weight with respect to the new weight function. In particular, we define the final weigh function as follows,
\[
w_{final} := w_{pl} \cdot n^l + w_{fks} , \text{for some constant } l>0.    
\]

Since both $w_{pl}$ and $w_{fks}$ are skew-symmetric, $w_{final}$ is also skew-symmetric. This settles the proof.
\end{proof}

\begin{corollary}
Deciding reachability in $\PSNS$ is in $\UL$.
\end{corollary}

\begin{proof}[of \cref{thm:pathmain}]
Given directed graph $G \in \PSNS$ on $n$ vertices and $2$ vertices $s$ and $t$, we iterate over all primes and for each of these primes, the weight function $w_{final}$ is computed according to \Cref{thm:maintheorem}. Thereby using \Cref{thm:ulalgorithm}, we verify if $G$ is min-unique with respect to $w_{final}$. If yes, we check if $t$ is reachable from $s$. If $G$ is not min-unique with respect to $w_{final}$, then we move to the next prime. The existence of such a prime with respect to which $G$ is min-unique is guaranteed by \Cref{lem:fks}. Hence, along a unique computation path, we get a \emph{Yes} or a \emph{No} answer depending on whether $s$ is reachable from $t$ or not respectively, while all other computation paths halt and reject. Since all the computation happens in logspace, this establishes the $\UL$ bound and settles the proof of \Cref{thm:pathmain}.
\end{proof}

Now, consider a graph $H$ having $O(n^k)$ many undirected cycles. We show that this implies that the number of directed cycles is bounded polynomially in $n$ for any directed version of $H$, say $G'$.   

%\todo{Remove this theorem}
\begin{theorem}
\label{thm:UnDirectedTheorem}
For an undirected $g$-genus graph $H$ having polynomially many surface non-separating cycles, there exists an $O(\log n)$-bit weight function $w': E \rightarrow \mathbb{Z}$ computable in logarithmic space that assigns non-zero weights to all the surface non-separating cycles in $H$. Moreover, for every graph $H \in \PSNS$, the minimum weight path in $H$ with respect to $w'$ is unique.
\end{theorem}

\begin{proof}
Let $t$ be the number of undirected simple SNSC in $H$ (each undirected cycle is a closed simple sequence of distinct vertices and edges). We create a directed graph $G'$ by replacing every undirected edge in $H$ with two directed edges in opposite directions. It is clear that each undirected simple cycle $C \in H$ produces exactly two distinct simple directed cycles in $G'$ - one following the edges of $C$ in one direction and the other following edges in the opposite direction. Let us denote one of those directions as \emph{clockwise} and the other as \emph{counterclockwise}. Then each undirected cycle contributes one clockwise directed simple cycle and one counterclockwise directed simple cycle. While considering clockwise orientation for every undirected cycle, the number of directed simple cycles oriented clockwise thus formed in $G'$ is $t$. Similarly, the number of counterclockwise oriented directed simple cycles in $G'$ also equals $t$, and the total number of directed cycles in $G'$ is $2t$ (polynomial in $n$, since $t = O(n^k)$).

Given the $g$-genus graph $G'$ having $2t$ many cycles, let $\mathcal{P}$ be the set of paths that form these cycles. We now apply \Cref{thm:maintheorem} on the directed graph $G'$ to get a $O(\log n)$-bit weight function $w'$ that assigns non-zero weights to all the SNSCs in $G'$. In the first stage, each edge is assigned $w'_{exp}$ with respect to which, each path in $\mathcal{P}$ get a different weight. In the second stage, the FKS Hashing scheme is used to compute an $O(\log n + \log t)$ bit weight function $w'_{fks}$ with respect to which only one among the many paths of the first stage gets the minimum weight value. Applying FKS Hashing Lemma is possible because the number of directed cycles is still polynomial in $t$, thereby also polynomial in $n$. Thus, with respect to an $O(\log n)$ bit weight function $w'_{final}$ created accordingly, $G'$ is min-unique. Combining the above argument with \Cref{prop:skewsymmetric}, if there exists a skew-symmetric weight function $w'_{final}$ that is path-isolating for $G'$, then for the underlying undirected graph $H$ of $G'$, there exists some $O(\log n)$-bit weight function $w'$, with respect to which, the minimum weight path in $H$ is unique. This concludes the proof of \Cref{thm:UnDirectedTheorem}.
\end{proof}

\section{Isolating bipartite perfect matchings in \texorpdfstring{$\PSNS$}{\PSNS}}
\label{Sec_perfectmatch}

\begin{definition} \cite{greenstheorem}
Given an undirected graph $H$, let $\mathcal{O}_H$ be the set of all directed graphs $H'$, such that the underlying undirected graph of $H'$ is $H$. For a class of undirected graphs $\mathcal{H}$, define $\mathcal{O}_\mathcal{H} = \bigcup_{H \in \mathcal{H}}  \mathcal{O}_H$.  
\end{definition}

\begin{theorem} \cite{greenstheorem}
\label{thm:Green_Theorem_PM}
Let $\mathcal{G}$ be a class of directed graphs and let $w$ be an edge weight function defined for every $G \in \mathcal{G}$ such that $w$ is skew-symmetric and for any cycle $C \in G$, $w(C) \neq 0$. Let $\mathcal{H}$ be a class of undirected bipartite graphs, such that $\mathcal{O}_{\mathcal{H}} \subseteq \mathcal{G}$. Then we can construct a weight function $w'$ in log-space, such that for every graph $H \in \mathcal{H}$, the minimum weight perfect matching in $H$ with respect to $w'$ is unique.    
\end{theorem}

\begin{theorem}
\label{thm:perfectmatch}
For the class of undirected bipartite graph $\mathcal{H}$ with polynomially many SNSCs, there exists an $O(\log n)$-bit weight function $w'_{pm}: E \rightarrow \mathbb{Z}$ computable in logarithmic space that assigns non-zero weights to all the surface non-separating cycles in $H$ such that with respect to $w'_{pm}$, for every graph $H' \in \mathcal{H}$, the minimum weight perfect matching is unique.
\end{theorem}

\begin{proof}
Let $\hat{H}$ be the directed version of the underlying undirected bipartite graph $H \in \mathcal{H}$ with $(L,R)$ being the bipartition and all the edges are directed from $L$ to $R$. For two perfect matchings $M_1$ and $M_2$ in $\hat{H}$, we define $E_{M_1 \square M_2} = (M_1 \cup M_2) \setminus (M_1 \cap M_2)$. The weight function $w_{pm}$ is the same as the weight function $w_{final}$ derived in the previous subsection for isolating minimum weight unique paths.

\begin{lemma}
\label{lem:Match_min_wt}
For a directed bipartite graph $\hat{H} \in \PSNS$, there exists at most one minimum weight perfect matching in $\hat{H}$ with respect to the weight function $w_{pm}$. 
\end{lemma}

\begin{proof}[of \cref{lem:Match_min_wt}]
Let us assume that $M_1$ and $M_2$ are two minimum weight perfect matching in $\hat{H}$ The set of edges in $E_{M_1 \square M_2}$ forms vertex-disjoint cycles, say $C_1, C_2, \ldots, C_k$, as they contain edges which are either in $M_1$ or in $M_2$ but not both. are either in $M_1$ or in $M_2$ but not both. Note that any cycle $C_i$ formed thus contain even number of edges alternating between $M_1$ and $M_2$ and all edges in this cycle are directed from $L$ to $R$.

Let $E_{i1}, E_{i2}$ be the set of edges in $M_1$ and $M_2$ respectively for a cycle $C_i$. We see that $w_{pm}(E_{j1}) = w_{pm}(E_{j2})$ for some cycle $C_j$. Otherwise, without loss of generality, if $w_{pm}(E_{j1}) > w_{pm}(E_{j2})$, then a new perfect matching $(M_1 \setminus E_{k1}) \cup E_{k2}$ can be formed having strictly lesser weight. This is a contradiction to the assumption that $M_1$ is a minimum weight perfect matching.

Now consider another graph $\hat{H'}$ which is same as $\hat{H}$ but the direction of edges in $M_2 \in \hat{H}$ are reversed. Let $M_1', M_2' \in \hat{H'}$ be the corresponding matchings to $M_1, M_2 \in \hat{H}$ such that the underlying undirected edges of $M_1$ and $M_2$ are the same as that of $M_1'$ and $M_2'$. Let $C_1', C_2', \ldots C_k'$ be the set of vertex disjoint cycles formed from the edges in $E_{M_1 \square M_2}$ and $E_{i1}', E_{i2}'$ be the edges in $M_1', M_2'$ respectively in a cycle $C_i'$.

We have seen that $w_{pm}(E_{i1}) = w_{pm}(E_{i2})$, for all $i \in [k]$. The set of edges $E_{i1}$ equals $E_{i1}'$ and the set of edges $E_{i2}$ equals the reversed set of edges $E_{i2}'$. Since $w_{pm}$ is skew-symmetric, combining the previous assertion that $w_{pm}(E_{j1}) = w_{pm}(E_{j2})$, we get,
\[w_{pm}(E_{i1}') = -w_{pm}(E_{i2}').\]
Therefore,
\[
w_{pm}(E_{i1}') + w_{pm}(E_{i2}') = 0, \text{and hence, } w_{pm}(C_i') = 0.
\]

It is known that $w_{pm}$, like $w_{final}$, assigns non-zero weights to all the SNSCs in $\hat{H}$. Therefore, $w_{pm}(C_i') \neq 0$ and this is a contradiction. Thus, there exists at most one minimum weight perfect matching in $\hat{H}$ with respect to $w_{pm}$. Note that, since the edges in $E_{i2}'$ are reversed they have a direction from $R$ to $L$ whereas edges in $E_{i1}'$ are from $L$ to $R$. Therefore, the cycles $C_1', C_2', \ldots C_k'$ that were formed are directed cycles in $\hat{H'}$. 
\end{proof}

We also know from the proof of \Cref{thm:UnDirectedTheorem} that if the number of SNSCs in the class of undirected bipartite graphs is polynomial in $n$, then the number of directed cycles in the corresponding directed version of the graphs, including cycles oriented both clockwise and anti-clockwise, is also bounded polynomially in $n$.

For the class of directed graphs $\hat{H} \in \PSNS$, there exists a skew-symmetric weight function $w_{pm}$, that gives non-zero weights to all the cycles in any graph $H \in \hat{H}$. Then using \Cref{thm:Green_Theorem_PM}, we can say that for the class of undirected bipartite graph $\mathcal{H}$, another weight function, say $w'_{pm}$, that can be constructed in logspace, such that for every $H' \in \mathcal{H}$, if there exists a matching, the minimum weight perfect matching in $H'$ with respect to $w'_{pm}$ is unique. Thus, we see that for an undirected $g$-genus bipartite graph $H' \in \mathcal{H}$, with respect to the weight function $w'_{pm}$, if there is a perfect matching in $H'$, the minimum weight perfect matching in $H'$ is unique. This finishes the proof of \Cref{thm:perfectmatch}.
\end{proof}

\begin{lemma} \cite{bddgenus1}
\label{lem:Perfect_Match_ARZ}
For any weighted graph $G$ assume that the minimum weight perfect matching in $G$ is unique and also for any subset of edges $E' \subseteq E$, the minimum weight perfect matching in $(G \setminus E')$ is also unique. Then deciding if $G$ has a perfect matching is in $\SPL$. \end{lemma}

\begin{corollary}
$\BPMD$ in $\PSNS$ is in $\SPL$.
\end{corollary}

\begin{proof}[of \cref{thm:matchingmain}]
We have the weight function $w'_{pm}$ from \cref{thm:perfectmatch}, according to which if there exists a matching in $H' \in \mathcal{H}$, then the minimum weight perfect matching in $H'$ is unique. Then by simply applying \Cref{lem:Perfect_Match_ARZ} for the weighted graph $H'$, the problem of deciding if $H'$ has a perfect matching lies in $\SPL$. This completes the proof of \Cref{thm:matchingmain}.
\end{proof}

\section{Reachability via logarithmic-width genus decompositions}
\label{Sec_lgnedge_crossing_UL}

Graphs embedded on large genus surfaces may have different types of interactions among all their handles. We study these interactions by looking at potentially large genus graphs with genus separating \emph{partition}, that divide the underlying surface on which the graph is embedded, into two sub-surfaces with only a small number of graph edges crossing the partition, referred here as \emph{crossing edges}.

The class of $g$-genus graphs admitting partitions with at most $O(\log g)$ crossing edges gives an intermediate class between low genus graphs and unrestricted high genus graphs. This restriction limits the amount of connections between subgraphs and different groups of handles, while allowing either side to have considerable topological complexity. The subgraphs on either side of the partition may individually have large genus, exponentially many paths, and exponentially many surface non-separating cycles. Nevertheless, the number of edges through which a path can move between the two subgraphs is $O(\log g)$. Thus, the graph can be topologically complicated within its individual regions but has a comparatively simple interaction structure between those regions.

The significance of this structural restriction becomes apparent in the context of path isolation. If the two subgraphs induced by a partition are individually min-unique with respect to positive edge weights, then the number of minimum weight paths between any pair of vertices in their union is bounded by the number of crossing edge subsets. Consequently, only polynomially many candidate minimum paths remain, allowing a polynomially bounded FKS hashing refinement to establish min-uniqueness across the partition.

This suggests a decomposition based approach to reachability in high genus graphs, in which the difficulty of isolating paths is governed not solely by the number of handles or surface non-separating cycles, but also by the interaction structure between subgraphs. In particular, logarithmic crossing edge width provides a natural structural parameter for investigating graph classes that may contain graphs of large genus while still admitting efficient path isolation procedures.

The $O(\log n)$ crossing edges consequently help us construct a path isolating weight assignment for this class of graphs and subsequently establish a $O(\log g \cdot \log n)$ unambiguous space bound on its reachability question. When two already isolated $k$-genus graphs are joined across $O(\log n)$ edges, the number of new surface non-separating cycles relevant to breaking ties between preliminary minimum paths is also polynomial.

\subsection{Genus Partitions and Crossing Edges}

In this section, we define \emph{genus partition} of a graph and its corresponding \emph{crossing edges} between the subgraphs of the decomposition of a $g$-genus directed graph embedded in a polygonal schema having $4g$ sides. Let $G=(V,E)$ be a directed graph embedded on an orientable surface $\Sigma$ by an embedding $\varphi$.

\begin{definition}[Genus Partition]
A genus partition of \(G\) is a separating simple closed curve \(\gamma\subseteq\Sigma\), disjoint from the vertices of \(G\), such that \(\gamma\) divides \(\Sigma\) into two connected sub-surfaces of genus \(g_1\) and \(g_2\), such that \(g = g_1 + g_2\).
\end{definition}

$\gamma$ intersects the embedding of every edge transversely at most once. The curve induces a vertex partition $V = V_1 \cup V_2$.

\begin{definition}[Crossing Edges]
Given a genus partition \(\gamma\), an edge \(e\in E\) is a crossing edge if its endpoints belong to different parts of the induced vertex partition. The set of crossing edges is
\[E_\times(\gamma) = E(V_1,V_2)\cup E(V_2,V_1)\] where \(E(V_i,V_j)\) denotes the set of directed edges from \(V_i\) to \(V_j\).
\end{definition}

The crossing edge width or the number of edges crossing $\gamma$ is $(\gamma)=|E_\times(\gamma)|$. We call a partition logarithmic-width if $r(\gamma) \le c\log n$ for a fixed constant $c>0$. A hierarchical genus partition is uniformly logarithmic-width if this condition holds at every internal node of its decomposition tree. This genus partition restriction is relevant when obtaining min-isolating weights for the two decomposing subgraphs of a graph while constructing the desired genus decomposition. Suppose, initially, that $g=2^d, d=\log_2 g$. The recursion has $d=\log_2 g$ merging levels.

Let $G=(V,E)$ be an $n$-vertex directed graph embedded on an orientable surface $\Sigma$ of genus $2k$. Let $\gamma$ be a genus-separating partition that divides the surface into 2 $k$-genus sub-surfaces and induces a vertex partition $V = V_1 \cup V_2$. Denote the induced subgraphs by $G_1 = G[V_1]$ and $G_2 = G[V_2]$. The set of crossing edges is $E_\times = \{(u,v) \in E\}$ such that $\{u \in V_1, v \in V_2\}$ or $\{u \in V_2, v \in V_1\}$. Suppose $|E_\times| \le c \cdot \log n$, for some constant $c>0$. For a positive integer edge-weight function \(w:E\to\mathbb N_{>0}\), define $\mathcal P_w(u,v) = \{P : P$ is a minimum-$w$-weight directed $u-v$ path$\}$. A graph is min-unique with respect to $w$ if $|\mathcal P_w(u,v)|\le1$, for every ordered pair $(u,v)$.

\subsection{Reachability bound and proof outline}

In this section, we study reachability in directed $g$-graphs embedded on surfaces that have an $O(\log n)$ crossing edges at every level of balanced binary decomposition of the surface $\Sigma$ in which the graph is embedded. The sub-surfaces containing subgraphs are decomposed recursively until each leaf corresponds to a $1$-genus piece. This decomposition-based bound is stated as \Cref{thm:crossmain} in Section~1.1 and proved below.

\subsection{Isolating paths in \texorpdfstring{$g$}{g}-genus graphs with \texorpdfstring{$O(\log n)$}{O(log n)} crossing edges}

We adapt the key lemma from \cite{kalampally} that adds $O(\log n)$ bits in each round for our case. Our modification replaces their path length doubling method with a genus doubling method. Instead of extending isolation from paths of length $\ell$ to $2\ell$, we extend it from two already isolated $k$-genus pieces to their $2k$-genus union.

We prove \Cref{thm:crossmain} in three parts. In the first part, we prove the \emph{Crossing Edges Merging Lemma} that given two min-unique subgraphs and $O(\log n)$ crossing edges between them, there exist only polynomially many competing minimum paths in the union of the subgraphs. In the next step, we prove the \emph{Genus Doubling Isolation Theorem} and show that by using only one $O(\log n)$-bit FKS refinement, we can get a minimum unique path at that level of merging subgraphs. Finally, in the last part, we construct the \emph{Recursive Level Isolation} by iterating over $O(\log g)$ levels of decomposition and establish the $O(\log g \cdot \log n)$ bound mentioned in \Cref{thm:crossmain}.

Let $G=(V,E)$ be a directed graph over $n$ vertices with a vertex partition $V$ and crossing edge set $E_\times$, such that $V = V_1 \cup V_2$ and $|E_\times| = r$. 

\begin{lemma} [Crossing Edges Merging Lemma]
\label{lem:CrossEdge_Merge}
Let $w:E \rightarrow \mathbb N_{>0}$ be a positive edge weight function such that both induced subgraphs $G[V_1]$ and $G[V_2]$ are min-unique with respect to $w$. Then, for every ordered pair of vertices \(u,v\in V\), $|\mathcal P_w(u,v)|\le2^r$. In particular, if \(r\le c\log_2 n\), then $|\mathcal P_w(u,v)|\le n^c$.
Consequently, the total number of minimum weight paths over all ordered vertex pairs is at most
$\left|\bigcup_{u,v\in V}\mathcal P_w(u,v)\right| \le n^{c+2}$.

\end{lemma}

\begin{proof}
Let $u,v\in V$. Since $w$ is positive, every minimum weight $u-v$ path is simple. Therefore, every such path traverses each crossing edge at most once. For any minimum weight path $P$, we define $S(P)$ as those crossing edges that are in path $P$, $S(P)=E(P)\cap E_\times$. We first establish the following claim.

\begin{claim}
\label{clm:min_wt_edge_order}
For any two minimum weight $u-v$ paths $P$ and $Q$, if $S(P) = S(Q)$ then $P = Q$.
\end{claim}

\begin{proof}[of \cref{clm:min_wt_edge_order}]
Suppose $S(P)=S(Q)=\{e_1,\ldots,e_q\}$. We show that the crossing edges occur in the same order along \(P\) and \(Q\). Let $d(x)=\operatorname{dist}_w(u,x)$ denote the minimum weight of a directed \(u\)-\(x\) path. Since every prefix of a minimum weight path is also minimum weight, if crossing edge \(e_i\) occurs before crossing edge \(e_j\) on a minimum weight path, then $d\bigl(\operatorname{tail}(e_i)\bigr) < d\bigl(\operatorname{tail}(e_j)\bigr)$, where $\operatorname{tail}(e_j)$ is the trailing vertex in the directed edge $e_j$. The inequality is strict because all edge weights are positive. Thus, the order of crossing edges on a minimum weight path is determined uniquely by the distances of their tail vertices from \(u\). Consequently, \(P\) and \(Q\) must traverse the crossing edges in the same order.

Say, $P = P_0e_1P_1e_2\cdots e_qP_q$ and $Q=Q_0e_1Q_1e_2\cdots e_qQ_q$. Since the crossing edges occur in the same order, \(P_i\) and \(Q_i\) have the same endpoints and lie entirely within the same induced subgraph \(G[V_1]\) or \(G[V_2]\). Moreover, each \(P_i\) and \(Q_i\) is a minimum weight path between its endpoints within that subgraph. Otherwise, replacing it with a lower weight path would produce a lower weight \(u\)-\(v\) walk, contradicting the minimality of \(P\) or \(Q\). Since both induced subgraphs are min-unique, $P_i=Q_i$ for every $i\in\{0,\ldots,q\}$. Therefore, $P = Q$. This establishes the claim.
\end{proof}

Since a simple path traverses each crossing edge at most once, the number of possible crossing edge subsets is bounded by $2^{O(\log g)}=g^{O(1)}$, which is polynomial in the input size for polynomially bounded genus. $E_\times$ has at most $2^r$ subsets. Thus, $|\mathcal P_w(u,v)| \le 2^r$. When \(r\le c\log_2 n\), $|\mathcal P_w(u,v)| \le 2^{c\log_2 n} = n^c$. Summing over all \(n^2\) ordered vertex pairs gives 
\[\left|\bigcup_{u,v\in V}\mathcal P_w(u,v)\right| \le n^{c+2} = O(n^{c+2})\]
\end{proof}

\begin{theorem} [Genus Doubling Isolation Theorem]
\label{thm:genus_double}
Let \(G=(V,E)\) be an \(n\)-vertex directed graph embedded on an orientable surface of genus $2k$, and let \(\gamma\) be a genus partition dividing the surface into two \(k\)-genus subsurfaces. Let \(G_1,G_2\) be the corresponding induced subgraphs, and let \(E_\times\) denote the crossing edge set satisfying $|E_\times|\le c\log_2 n$, for a fixed constant \(c>0\). Suppose there exist positive integer edge-weight functions $w_1:E(G_1)\longrightarrow\mathbb N_{>0}$ and $w_2:E(G_2)\longrightarrow\mathbb N_{>0}$ such that both \(G_1\) and \(G_2\) are min-unique with respect to \(w_1\) and \(w_2\), respectively. Assume further that $\max\{w_1(e),w_2(e)\}<2^b$ for some \(b\ge1\).
Then there exists a positive integer edge-weight function 
$W:E(G)\longrightarrow\mathbb N_{>0}$ such that : 
\begin{enumerate}
    \item \(G\) is min-unique with respect to \(W\). 
    \item  Every edge weight \(W(e)\) can be represented using at most $b+O(\log n)$ bits.
    \item \(W(e)\) is computable in space \(O(b+\log n)\). In particular, it is logspace computable when \(b=O(\log n)\) and the child weights are logspace computable.
\end{enumerate}
\end{theorem}

\begin{proof}
We construct the desired weight function in three steps. In the first step, we construct the preliminary weight function $w : E(G) \rightarrow\mathbb N_{>0}$ as follows : 
\[
w(e)= \begin{cases}
w_1(e),&e\in E(G_1),\\[2mm]
w_2(e),&e\in E(G_2),\\[2mm]
1,&e\in E_\times.
\end{cases}
\]

By assumption, $G_1$ and $G_2$ are min-unique under the restrictions of $w$. Although $G$ need not be min-unique under $w$, \Cref{lem:CrossEdge_Merge} implies that
\[
|\mathcal P_w(u,v)| \le 2^{|E_\times|} \le n^c, \text{ for every $u,v \in V$.}
\]

Let $\mathcal S = \bigcup_{u,v\in V}\mathcal P_w(u,v)$. Then $|\mathcal S| \le n^{c+2}$. Thus, $w$ leaves only polynomially many minimum weight paths across all vertex pairs.

In the next step, we apply FKS hashing on these polynomially many min-weight paths. Let $E(G) = \{e_1,e_2,\ldots,e_m\}$ be a fixed ordering of the directed edges. Define an exponential edge-weight function $w_{\exp}$ as $w_{\exp}(e_i)=2^{i-1}$.
For a simple path \(P\), define
\[
w_{\exp}(P)=\sum_{e_i\in P}2^{i-1}.
\]
Because the paths in $\mathcal S$ are simple, distinct paths have distinct sets of directed edges. Therefore, if $P\ne Q$, then $w_{\exp}(P)\ne w_{\exp}(Q)$. Each value of $w_{\exp}(P)$ has at most $(m+1)$ bits. Since \(m\le n^2\) for a simple directed graph, and \(|\mathcal S|\le n^{c+2}\), \cref{lem:fks} guarantees the existence of a prime \(p\) having \(O(\log n)\) bits such that
\[
P\ne Q,\quad P,Q\in\mathcal S
\implies
w_{\exp}(P)\not\equiv
w_{\exp}(Q)\pmod p.
\]

Define
\[
h_p(e_i)=2^{i-1}\bmod p.
\]

Therefore, for any two distinct paths $P,Q \in \mathcal S$,
\[
\begin{aligned}
h_p(P)-h_p(Q)
&\equiv
w_{\exp}(P)-w_{\exp}(Q)
\pmod p\\
&\not\equiv0\pmod p.
\end{aligned}
\]

%/Here \(h_p(P)\) is the ordinary integer sum of the hashed edge weights. In particular,
%/\[
%/h_p(P)\ne h_p(Q).
%/\]

Thus, \(h_p\) distinguishes every pair of distinct paths in \(\mathcal S\). While combining the preliminary weight with the FKS weight, we choose $B$ of some large value so that the component of the weight from the previous round gets more priority than the FKS component. Say $B=n(p-1)+1$. Then, we define the final edge-weight function as,
\[
W(e)=B\,w(e)+h_p(e).
\]

Now we show that $G$ is min-unique under $W$. Fix any two vertices $u,v$, and let $P$ and $Q$ be two simple directed $u-v$ paths. Since a simple path has at most \(n-1\) edges and the FKS weight is less than the prime $p$, $0\le h_p(e)<p$, we have
\[
0\le h_p(P), h_p(Q)\le(n-1)(p-1)<B.
\]

Therefore, when $w(P) < w(Q)$, the integrality of $w$ gives $w(Q) - w(P) \ge 1$.

Hence,
\[
\begin{aligned}
W(Q)-W(P)
&=
B(w(Q)-w(P))
+h_p(Q)-h_p(P)\\
&\ge B-(n-1)(p-1)\\
&>0.
\end{aligned}
\]

Conversely, if $w(P)<w(Q)$, then $W(P)<W(Q)$. In particular, every minimum $W$-weight $u-v$) path must also be a minimum $w$-weight $u-v$ path. Now suppose, towards a contradiction, that \(P,Q\) are distinct minimum $W$-weight $u-v$ paths. Both paths must belong to $\mathcal P_w(u,v)\subseteq\mathcal S$. Therefore, $w(P) = w(Q)$. Because $P\ne Q$ and $P,Q \in \mathcal S$, $h_p(P)\ne h_p(Q)$. Hence,

\[
\begin{aligned}
W(P)-W(Q)
&= B(w(P)-w(Q)) + h_p(P) - h_p(Q) \\
&= h_p(P)-h_p(Q) \\
&\ne 0.    
\end{aligned}
\]

This contradicts the assumption that \(P,Q\) have equal minimum weight $W$. Thus, for every $u,v$, $|\mathcal P_W(u,v)| \le 1$. Therefore $G$ is min-unique with respect to $W$. Since \(p=n^{O(1)}\), $B=n^{O(1)}$. Also for every edge, $w(e) < 2^b$. Therefore, $W(e) < B\cdot2^b+p$, hence $W(e)$ requires $O \bigl(b+O(\log n) \bigr)$ bits. Given $p$ and the child weights, the function $W$ can be evaluated in $O(b+\log n)$ space. This finishes the proof of \Cref{thm:genus_double}.
\end{proof}

We observe that after the subgraphs have been isolated, complexity of merging the paths depend on the cut size over the partition and not directly on the total genus. Now we combine the weights over all $\log g$ stages. We extend \Cref{thm:genus_double} recursively to obtain a min-isolating weight function for directed $g$-genus graphs. The construction uses \Cref{lem:CrossEdge_Merge} and a single FKS hashing at each level by sharing the same prime for merging at that level.

\begin{corollary}[Recursive Level Isolation]
\label{cor:recgenus}
Let \(G=(V,E)\) be an \(n\)-vertex directed graph embedded on an orientable surface of genus \(g\). Suppose \(G\) has a binary genus logspace decomposition $T$ of depth $d=O(\log g)$, where every internal node \(X\) has two children \(X_1,X_2\), with $V_X = V_{X_1} \cup V_{X_2}$, and crossing edge set satisfying $|E_\times(X)| \le c \cdot \log n$ for some fixed constant $c$. If each leaf graph has a positive, logspace computable, $O(\log n)$-bit min-isolating weight function $W_0$, then $G$ admits a min-isolating weight function using $O(\log g \cdot \log n)$ bits per edge.
\end{corollary}

\begin{proof}[of \cref{thm:crossmain}]
Suppose every subgraph at level $j-1$ is min-unique under $W_{j-1}$. Let the nodes at level $j$ be $X_1, X_2, \ldots, X_{r_j}$. For each \(X_i\), define the preliminary weight $f_j$ by retaining $W_{j-1}$ on the two child subgraphs and assigning weight $1$ to their crossing edges.

Let
\[
\mathcal S_{X_i}
=
\bigcup_{u,v\in V_{X_i}}
\mathcal P_{f_j}(u,v).
\]

By \Cref{lem:CrossEdge_Merge}, $|\mathcal P_{f_j}(u,v)| \le 2^{|E_\times(X_i)|} \le n^c$. Therefore, $|\mathcal S_{X_i}| \le |V_{X_i}|^2n^c.$ Since the vertex sets at level \(j\) are disjoint,
\[
\sum_{i=1}^{r_j}|V_{X_i}|^2\le n^2.
\] 

Consequently,
\[
\left|\bigcup_{i=1}^{r_j}\mathcal S_{X_i}\right|
\le n^{c+2}.
\]

Thus, there are polynomially many preliminary minimum paths across all merges at level $j$. Index the edges of \(G\) as \(e_1,\ldots,e_m\). By the FKS Hashing Lemma, there exists a polynomially bounded prime $p_j$ with respect to which $h_j(e_i) = 2^{i-1}\bmod p_j$ distinguishes all distinct nontrivial paths in this collection. %/Zero-length paths can be omitted, since they do not cause a failure of min-uniqueness.%/ 
Hence, one \(O(\log n)\)-bit prime $p_j$ suffices for every merge at level $j$. 

In the next step, we construct the recursive weight assignment. For every node at level \(j\), define
\[
f_j(e)=
\begin{cases}
W_{j-1}(e),&e\text{ belongs to a child subgraph}, \\
1,&e\in E_\times(X).
\end{cases}
\]

Choose a polynomially bounded integer $B$ satisfying $B > n(p_j-1)$ for every level $j$. The same $B$ can be used throughout the construction. Let us define the weight function for an intermediate step $j$, $W_j(e)=Bf_j(e)+h_j(e)$.

As shown in \cref{thm:genus_double}, factor $B$ ensures that $f_j$ weight takes precedence over the hash contribution. Thus, every minimum $W_j$ path is a minimum $f_j$ path. Since $h_j$ distinguishes all distinct minimum $f_j$ paths, \(W_j\) makes every $j$th subgraph min-unique. Applying induction, $W_1,W_2,\ldots,W_d$ isolate all subgraphs at their respective levels. Thus, $G$ is min-unique with respect to $W_d$.

In the last step, we calculate the size of the final weight. Let $b_j$ denote the maximum number of bits required to represent an edge weight under $W_j$. Since $B$ and $p_j$ are polynomially bounded, we have $b_j \le b_{j-1}+O(\log n)$. The leaf subgraph requires $b_0 = O(\log n)$ bit weights. Therefore, $b_j = O\bigl( (j+1) \cdot \log n \bigr)$. Since the decomposition has depth of $O(\log g)$, the final weight requires a total of $b_d = O\bigl((1+\log g) \cdot \log n \bigr)$ bits. Thus, each genus doubling stage adds only $O(\log n)$ bits. This completes the proof of \Cref{thm:crossmain}.
\end{proof}

Moreover, directed reachability in $G$ can be decided by a polynomial time unambiguous algorithm using $O(\log g \cdot \log n)$ space. At each level $j$, we enumerate the candidate primes \(p_j\) in increasing order and use the min-uniqueness testing algorithm in \cite{kalampally} to select the prime making every level $j$ subgraph min-unique.

\Cref{thm:genus_double} provides an unambiguous min-uniqueness test for a $b$ bit weight function in $O(b + \log n)$ space. Storing the selected primes $p_1,p_2,\ldots,p_d,$ require $O(d \cdot \log n)$ bits altogether. Given these primes, the final edge weights can be evaluated in $O \bigl((d+1) \cdot \log n \bigr)$ space. Hence, the total space required for weight construction and min-uniqueness testing is $O \bigl((d+1) \cdot \log n \bigr) = O(\log g \cdot \log n)$. Applying the weighted min-unique reachability algorithm gives an unambiguous algorithm using the same space bound. Therefore, reachability in graphs with $O(\log g)$ crossing edges $\in \operatorname{USPACE} \bigl( O(\log g \cdot \log n)\bigr)$. This establishes the space bound of \Cref{thm:crossmain}.

%\begin{theorem}
%This is a sample theorem. The run-in heading is set in bold, while the following text appears in italics. Definitions, lemmas, propositions, and corollaries are styled the same way.
%\end{theorem}
%
% the environments 'definition', 'lemma', 'proposition', 'corollary',
% 'remark', and 'example' are defined in the LLNCS documentclass as well.
%
%\begin{proof}
%Proofs, examples, and remarks have the initial word in italics, while the following text appears in normal font.
%\end{proof}
%For citations of references, we prefer the use of square brackets and consecutive numbers. Citations using labels or the author/year convention are also acceptable. The following bibliography provides a sample reference list with entries for journal articles~\cite{ref_article1}, an LNCS chapter~\cite{ref_lncs1}, a book~\cite{ref_book1}, proceedings without editors~\cite{ref_proc1}, and a homepage~\cite{ref_url1}. Multiple citations are grouped
%\cite{ref_article1,ref_lncs1,ref_book1},
%\cite{ref_article1,ref_book1,ref_proc1,ref_url1}.

\section{AI Disclosure}We used GPT 6.1 Plus to assist with proof ideas, writing parts of the paper and reviewing drafts and statements of the work. The tool materially affected \cref{Sec_lgnedge_crossing_UL}. We verified the correctness and originality of all content, including references.

\acknowledgements
This work was supported in part by the PM ECRG Grant, India (Grant No. \path{ANRF/ECRG/2024/004480/ENS}), and the Faculty Initiation Grant (No. 101027), IIT Roorkee, awarded to Chetan Gupta.

% \subsubsection{\discintname}
% %It is now necessary to declare any competing interests or to specifically state that the authors have no competing interests. Please place the statement with a bold run-in heading in small font size beneath the (optional) acknowledgments\footnote{If EquinOCS, our proceedings submission system, is used, then the disclaimer can be provided directly in the system.}, for example: The authors have no competing interests to declare that are relevant to the content of this article. Or: Author A has received research grants from Company W. Author B has received a speaker honorarium from Company X and owns stock in Company Y. Author C is a member of committee Z.

\begin{comment}

\end{comment}

%\nocite{*}
\bibliographystyle{abbrvnat}
% use the following instead if you encounter problems 
%\bibliographystyle{alpha}
\bibliography{mybibliography}
\label{sec:biblio}

\end{document}